\documentclass[11pt]{article}
\usepackage{amsmath,amssymb,amscd}
\usepackage{cite}
\usepackage[usenames]{color}

\usepackage{graphicx}
\usepackage[mathscr]{eucal}
\usepackage{color}
\large\normalsize

\title{Dimensionality-induced critical phase transition in stochastic Lotka-Volterra equation: From statistical averaging to systemic tipping point}
\author{Xiu-deng Zheng$^1$, Cong Li$^2$, Hui Zhang$^3$, Hui-jie Qiao$^1$\footnote{Author for correspondence, and Email: qiaohj@ioz.ac.cn}, Shao-peng Wang$^4$  and Yi Tao$^{1,2}$\footnote{Author for correspondence, and Email: yitao@ioz.ac.cn}
\\
$^1$State Key Laboratory of Animal Biodiversity Conservation \\ and Integrated Pest Management, Institute of Zoology, \\
Chinese Academy of Sciences, Beijing 100101, China \\
$^2$School of Life Science and Technology,\\ Northwestern Polytechnical University, Xi'an, Shaanxi 710072,  China\\
$^3$School of Mathematics and Statistics, \\
Northwestern Polytechnical University, Xi'an, Shaanxi 710072, China\\
$^4$State Key Laboratory of Vegetation Structure, Function and Construction (VegLab), \\ Institute of Ecology, College of Urban and Environmental
Science, \\ Peking University, Beijing 100871,China.
}
\date{}

\begin{document}
\maketitle

\newpage

\emph{\textbf{Abstract.}} By analyzing a stochastic Lotka-Volterra (LV) equation, we show that the underlying logic behind the diversity-stability debate in ecology can be framed as a dimensionality-induced critical phase transition, that is, a transition separating the regime dominated by statistical averaging and a systemic tipping point. This phase transition framework unifies the opposing ecological predictions in the diversity-stability debate and sets an intrinsic diversity ceiling for stochastic ecological communities.

\newpage

\emph{\textbf{Introduction.}} In the past more than fifty years, the relationship between biodiversity and ecosystem stability has remained a central debate in ecology \cite{macarthur1955,may1972,may1973,tilman1994,mccann2000, loreau2001,thebault2005,tilman2006,ives2007,allesina2012}, where the diversity is defined as the number of species in a community in general. A lot of ecological field observations, as well as experimental and statistical ecological studies, claim that the more complex a community is, the greater its stability (where the community stability is generally measured by its temporal variability, i.e. the coefficient of variation, $CV$) \cite{tilman1994,thebault2005, tilman2006,ives2007, downing2020, liang2025}. The stabilizing effects of biodiversity is attributed to the statistical averaging of asynchronous species responses to environmental fluctuations \cite{mccann2000, loreau2001,thebault2005,zhao2022}. However, as early as 1972, by the analysis for a random community matrix stability model \cite{may1972,may1973}, theoretical ecologist Robert May found that diversity tends to destabilize the dynamic stability of the community - known as the so-called \emph{diversity-stability debate} \cite{mccann2000, loreau2001}.

So far, the basic consensus in academia on the diversity-stability debate is that the debate is really just due to the difference between two ways of measuring community stability: one is the temporal variability of community and the other dynamic stability of community \cite{mccann2000}. However, this consensus didn't really get to the underlying logic behind the diversity-stability debate.  Actually, from a nonlinear dynamics perspective, the diversity-stability relationship should involve mainly how changes in the dimension of a stochastic dynamical system affect its steady-state statistical properties (e.g. the temporal variability) and the critical phase transitions in stochastic stability \cite{mao1994, mao2007, arnoldi2019}.  So, using a unified model ecosystem to reveal how the changes in diversity affects both the community's steady-state statistics and its stochastic stability, or more precisely, how changes in diversity influence the intrinsic link between them, should be key to resolving the diversity-stability debate.

We here consider a competitive system in a stochastic environment, described by a simple Lotka-Volterra (LV) equation \cite{hofbauer1998}. For simplicity, we also assume that only the intraspecific and interspecific interactions are influenced by the environmental stochastic fluctuations \cite{may1972,dong2015,arnoldi2019,chen2025}.  We found that increasing diversity generally tends to reduce the community's temporal variability, which is fully consistent with the results of many experimental and theoretical studies \cite{tilman1994,loreau2001,tilman2006,thibaut2013,wang2019}. However, once the diversity exceeds a critical threshold, the community will lose its stochastic stability - the capacity to return toward equilibrium under persistent environmental stochastic fluctuations, which aligns with theoretical results including May \cite{may1972}. Therefore, the changes in diversity create a non-monotonic, two-phase regime: a stable phase when the diversity is below the critical threshold, in which the community is not only stochastically stable but also increasing diversity suppresses community fluctuations; following by an unstable phase once the diversity exceeds the critical threshold, making the community extremely vulnerable to stochastic perturbations. This finding clearly demonstrate that, within the diversity-stability controversy, the two distinct ways of measuring community stability and their corresponding conclusions are not fundamentally contradictory - at least from the stochastic dynamics perspective, that is, they separately describe how the diversity variation shapes community stochastic dynamics across two distinct dynamic regimes.

Our results recast the diversity-stability relationship as a critical phase transition. This framework can not only unify opposing ecological predictions in diversity-stability debate, but also reveals how environmental stochasticity limits a  community's critical diversity threshold.

\emph{\textbf{A basic model.}} Consider a $S$-species competitive system in a stochastic environment, which can be described by a stochastic Lotka-Volterra (LV) equation (called as the Langevin equation \cite{risken1992,kampen1997}). Similar to May's model \cite{may1972}, a simple stochastic LV equation can be given by
\begin{align}
\frac{dn_i}{dt} &= n_i \Bigg ( \alpha -\beta_i (t) n_i -\sum_{j \ne i} \mu(t) \gamma_{ij}(t) n_j \Bigg )
\end{align}
for $i=1,2,\cdots,S$, in which $n_i$ denotes the size (or biomass) of species $i$; $\beta_i(t)$ denotes the intraspecific interaction strength of species $i$, defined as $\beta_i(t)=\beta + \xi_i(t)$, where $\beta$ is a positive constant and $\xi_i(t)$ is a white noise with $\left<\xi_i(t) \right > =0$ and $\left < \xi_i(t) \xi_i(t') \right > =2 \mathcal{D} \delta (t-t')$ \cite{risken1992,kampen1997}; $\gamma_{ij}(t)$ denotes the effect of interspecific interaction between species $i$ and species $j$ on species $i$, defined as $\gamma_{ij}(t)=\gamma +\zeta_{ij}(t)$, where $\gamma$ is a non-negative constant and $\zeta_{ij}(t)$ is a white noise with $\left <\zeta_{ij}(t) \right >=0$ and $\left < \zeta_{ij}(t) \zeta_{ij}(t') \right > =2 \mathcal{Q} \delta(t-t')$ for all possible $i \ne j$; and $\mu(t)$ is a random variable, defined as $\mu(t)=1$ with probability $C$ and $\mu(t)=0$ with probability $1-C$, where $C$ is used to measure the connectance among different species in this system \cite{may1972, allesina2012}. This implies that all species share the same expected parameter values. In addition, for simplicity, we also assume that all random variables in our model are independent of each other, or, more precisely, all $\xi_i(t)$ are independent and identically distributed, all $\zeta_{ij}$ are independent and identically distributed, and $\xi_i(t)$ and $\zeta_{kl}(t)$ and independent of each other for all possible $i$ and $(k,l)$.

Note that the mean and variance of the term $\mu(t) \gamma_{ij}(t)$ in Eq. (1) can be given by $C \gamma$ and $C \big [ 2 \mathcal{Q} +(1-C) \gamma^2 \big ]$, respectively, for all possible $i \ne j$. Thus, for large $C$ being close to $1$, the term $\mu(t) \gamma_{ij}(t)$ can also be approximated as a Gaussian random variable, defined as $\mu(t) \gamma_{ij}(t) = C \gamma + \upsilon_{ij}(t)$ with $\left< \upsilon_{ij}(t) \right > =0$ and $\left < \upsilon_{ij}(t) \upsilon_{ij}(t') \right >= 2 \mathcal{H} \delta(t-t')$, where $\mathcal{H}=(C/2) \big [ 2 \mathcal{Q} +(1-C) \gamma^2 \big ]$.  In this study, for convenience, we consider only the case with large $C$.  Therefore, Eq. (1) can be re-expressed as
\begin{align}
\frac{dn_i}{dt} &= n_i \Bigg ( \alpha - \beta n_i - \sum_{j \ne i} C \gamma n_j \Bigg ) -n_i \Bigg ( \xi_i(t) n_i + \sum_{j \ne i} \upsilon_{ij}(t) n_j \Bigg )
\end{align}
for $i=1,2, \cdots, S$. We can see that
\begin{align}
\frac{dn_i}{dt} &= n_i \Bigg ( \alpha - \beta n_i - \sum_{j \ne i} C \gamma n_j \Bigg )
\end{align}
for $i=1,2,\cdots,S$ represents a simple deterministic $S$-species LV equation. This equation has a unique positive equilibrium point $\textbf{n}^*=(n_1^*, n_2^*, \cdots, n_S^*)$ with $n_i^*\equiv n^*=\alpha \big / \big (\beta+(S-1) C \gamma \big)$ for $i=1,2,\cdots,S$, and $\textbf{n}^*$ is globally asymptotically stable if and only if $\beta-C \gamma >0$ (the proof is shown in Supporting Information). Obviously, this equation is not really realistic because it allows any number of species to coexist stably. However, by analyzing the stochastic dynamic properties near the equilibrium $\mathbf{n}^*$, we can reveal how the changes in diversity (i.e. dimensional variation) will affect the system's steady-state statistics and stochastic stability. So, in this study, we only focus our attention on the stochastic dynamic properties of the system near $\mathbf{n}^*$.

\emph{\textbf{Steady-state statistics of the system.}} When the system state is near $\mathbf{n}^*$, Eq. (2) can be approximated as
\begin{align}
\frac{dx_i}{dt} &= -n_i^* \Bigg ( \beta x_i + \sum_{j \ne i} C \gamma x_j \Bigg ) - n_i^* \Bigg ( \xi_i(t) n_i^* + \sum_{j \ne i} \upsilon_{ij}(t) n_j^* \Bigg )
\end{align}
for $i=1,2, \cdots, S$, where $x_i=n_i-n_i^*$. Let $p(\mathbf{x},t)$ denotes the probability density distribution that the system state equals exactly $\mathbf{x}$. Then, the Fokker-Planck equation corresponding to the above equation can be given by
\begin{align}
\frac{\partial p(\mathbf{x},t)}{\partial t} &= \sum_{i=1}^S \Bigg [ \frac{\partial}{\partial x_i} n_i^* \Big (\beta x_i + \sum_{j \ne i} C \gamma x_j \Big ) p(\mathbf{x},t) \nonumber \\
& \ \ \ \ \ \ \ \ \ \ \ \ + \frac{\partial^2}{\partial x_i^2} (n_i^*)^2 \Big (\mathcal{D} (n_i^*)^2 +\sum_{j \ne i} \mathcal{H} (n_j^*)^2 \Big ) p(\mathbf{x},t) \Bigg ]
\end{align}
\cite{risken1992}.  Also, if $\mathbf{x}=\mathbf{0}$ (i.e. the point $\mathbf{n}^*$) is stochastically stable, then there must be a unique stationary distribution near $\mathbf{x}=\mathbf{0}$, that is, $p(\mathbf{x})= \lim \limits_{t \rightarrow \infty} p(\mathbf{x},t)$ \cite{risken1992, kampen1997}.

Note that the definition of the stochastic LV equation in Eq. (1) implies that all species have the same expected size and the same variance, and the covariance between different species is also the same. Thus, based on the Fokker-Planck equation in Eq. (5), and from the steady solutions of $d\left<x_i \right> \big / dt$ and $d \left<x_i^2 \right> \big / dt$ for $i=1,2,\cdots,S$, and that of $d \left <x_i x_j \right > \big / dt$ for all possible $i \ne j$, the expected size of each species is $n^*$, and the variance of each species, denoted by $\sigma^2$, and the covariance between different species, denoted by $Cov$, can be given by
\begin{align}
& \ \sigma^2 = \frac{\alpha^3 \big (\mathcal{D}+(S-1) \mathcal{H} \big ) \big ( \beta + (S-2) C \gamma \big )}{( \beta -C \gamma) \big ( \beta + (S-1) C \gamma  \big )^4} \ , \nonumber \\
& \ Cov = - \frac{\alpha^3 \big (\mathcal{D} +(S-1) \mathcal{H} \big ) C \gamma}{(\beta -C \gamma) \big ( \beta + (S-1) C \gamma \big )^4} < 0 \ ,
\end{align}
respectively (see Supporting Information). Therefore, the variance of the total system size can be expressed as
\begin{align}
\sigma_{total}^2 &= S \sigma^2 +S(S-1) Cov \nonumber \\
&= \frac{S \alpha^3 \big (\mathcal{D}+(S-1) \mathcal{H} \big )}{\big (\beta +(S-1) C \gamma \big )^4} \ ,
\end{align}
and the coefficient of variation ($CV$) can be given by
\begin{align}
CV &= \sqrt{\frac{\sigma_{total}^2}{(Sn^*)^2}} = \sqrt{\frac{\alpha \big (\mathcal{D}+(S-1) \mathcal{H} \big )}{S \big (\beta +(S-1) C \gamma \big )^2}}  \ ,
\end{align}
also known as the \emph{temporal variability} of community \cite{tilman1994,thebault2005, tilman2006,ives2007, downing2020, liang2025} and its reciprocal has been widely used to measure the degree of of community stability \cite{tilman1994,thebault2005, tilman2006,ives2007, downing2020, liang2025}.

The above results implies that if $\mathcal{D} \ge \mathcal{H}$, then $\partial CV \big / \partial S < 0$, or more precisely, increasing diversity will definitely reduce the community's temporal variability (see Fig. 1a). This result is undoubtedly consistent with the current mainstream views \cite{tilman1994,ives2002,caldeira2005,Valencia2020}. However, if $\mathcal{D}<\mathcal{H}$, then there exists a $\tilde{S}$, which is given by
\begin{align}
\tilde{S} &= \frac{3(\mathcal{H}-\mathcal{D}) + \sqrt{9(\mathcal{H}-\mathcal{D})^2 (C\gamma)^2+8\mathcal{H}(\mathcal{H}-\mathcal{D}) (\beta -C\gamma) C \gamma}}{4 \mathcal{H} C \gamma} \ ,
\end{align}
such that the $CV$ will increase with the increase of $S$ if $S< \tilde{S}$, and the $CV$ will decrease with the increase of $S$ if $S> \tilde{S}$ (see Fig. 1b). This result strongly suggests that when the diversity level of a community is very low, increasing diversity may also lead to an increase in its temporal variability.

\begin{figure}[htbp]
    \centering
    \includegraphics[width=13.6cm,height=5.22cm]{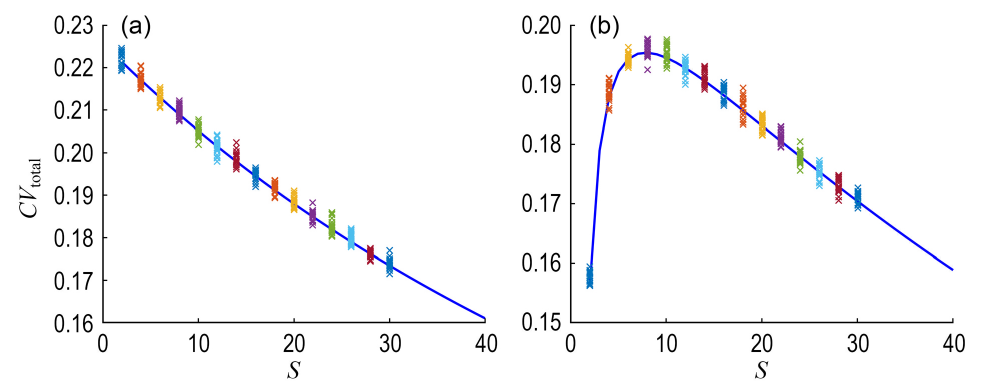}
    \caption{\emph{\textbf{The temporal variability ($CV$) of the community versus the species diversity ($S$).}} The solid curves show the theoretical predictions, and the crosses represent the stochastic simulation results. The parameters $\alpha$, $\beta$, $\gamma$ and $C$ are taken as $\alpha=10$, $\beta=1$, $\gamma=0.01$ and $C=1$, respectively. The simulations are repeated $20$ times for each $S$, in which $\mathcal{D}=\mathcal{Q}=0.005$ in panel (\textbf{a}), and $\mathcal{D}=0$ and $\mathcal{Q}=0.005$ in panel (\textbf{b}).}
    \label{Fig1}
\end{figure}

\emph{\textbf{Stochastic stability of the system.}} We now consider the effect of changes in diversity on the critical phase transition of stochastic stability of the point $\mathbf{n}^*$.  Similar to May's model \cite{may1972} (see also \cite{ives2002, caldeira2005, Valencia2020}), for Eq. (1), when the system state is in an infinitesimal neighborhood of the point $\mathbf{n}^*$, the community matrix can be given by a random matrix $\mathbf{A}(t)$ with entries $a_{ii}(t)= -n^* \big (\beta + \xi_i(t) \big )$ for $i=1,2, \cdots S$ and $a_{ij}(t)=-n^* \big (C \gamma +\upsilon_{ij}(t) \big )$ for all $i \ne j$. So, the It$\hat{\text{o}}$ stochastic differential equation (SDE) corresponding to $\mathbf{A}(t)$ can be given by
\begin{align}
d \mathbf{x}(t) &= \Bigg [ \tilde{\mathbf{A}} dt + \sum_{i=1}^S \sum_{j=1}^S \mathbf{B}^{(i,j)} d \omega_{ij}(t) \Bigg ] \mathbf{x}(t) \ ,
\end{align}
where $x_i=n_i-n^*$ for $i=1,2,\cdots,S$, $\tilde{\mathbf{A}}$ is a $S \times S$ matrix with entries $\tilde{a}_{ii}=-n^* \beta$ and $\tilde{a}_{ij}=-n^* C \gamma$ for $i \ne j$, $\mathbf{B}^{(i,j)}$ with $i \ne j$ is a $S \times S$ matrix with entries $b_{ij}^{(i,j)} = -n^* \sqrt{2\mathcal{H}}$ and $b_{kl}^{(i,j)} = 0$ for all $(k,l) \ne (i,j)$, and $\mathbf{B}^{(ii)}$ is a $S \times S$ matrix with entries $b_{ii}^{(i,i)} = -n^* \sqrt{2\mathcal{D}}$ and $b_{kl}^{(i,i)} = 0$ for all $(k,l) \ne (i,i)$ \cite{mao1994, mao2007}.

From the stochastic stability theory of SDE \cite{mao1994, mao2007},  for $S > 1$,  the zero solution of Eq. (10), $\mathbf{x}=\mathbf{0}$ (i.e. the point $\mathbf{n}^*$), is said to be stochastically stable, or mean-square exponentially stable, if and only if
\begin{align}
& \ n^* \big (\mathcal{D} +(S-1) \mathcal{H} \big ) < \beta -C \gamma \nonumber \\
\Rightarrow & \ \ \ \frac{\alpha \big (\mathcal{D} +(S-1) \mathcal{H} \big )}{\beta +(S-1) C \gamma} < \beta -C \gamma
\end{align}
(the proof is shown in Supporting Information). As a special case, if take $\alpha=\beta=1$, $\gamma=0$, and $\mathcal{D}=0$, then $\mathbf{x}=\mathbf{0}$ is stochastically stable if and only if $(S-1) \mathcal{Q}C<1$. We can see that this result exactly matches May's results \cite{may1972}. Furthermore, for $S=1$, the zero solution of Eq. (10) is stochastically stable if and only if $\alpha \mathcal{D} < \beta^2$ (the proof is shown in Supporting Information).

The criterion for the stochastic stability in Eq. (11) implies that if $\mathcal{H}/C \gamma > \mathcal{D}/\beta$, that is, the relative fluctuation intensity of interspecific interactions is greater than that of intraspecific interactions, and $\alpha \mathcal{D} / \beta < \beta -C \gamma < \alpha \mathcal{H} / C \gamma$, then there must exist a critical diversity threshold, denoted by $S_c$, which is given by
\begin{align}
S_c &= \frac{(\beta -C\gamma)^2 +\alpha ( \mathcal{H} - \mathcal{D})}{\alpha \mathcal{H} -(\beta -C \gamma) C \gamma} \ ,
\end{align}
such that if $S > S_c$, then $\mathbf{x}=\mathbf{0}$ loses its stochastic stability and the stationary distribution near $\mathbf{x}=\mathbf{0}$ also vanishes (see Fig. 2a).

For different values of $S$, the results of stochastic simulations show clearly the probability that at least one species disappears in the system in different time periods (see Fig. 2b). We can see that for the cases with $S>S_c$, this probability quickly approaches $1$ as the time period gets longer. Obviously, these stochastic simulation results match the theoretical predictions.

\begin{figure}[htbp]
    \begin{center}
    \includegraphics[width=13.6cm,height=5.6cm]{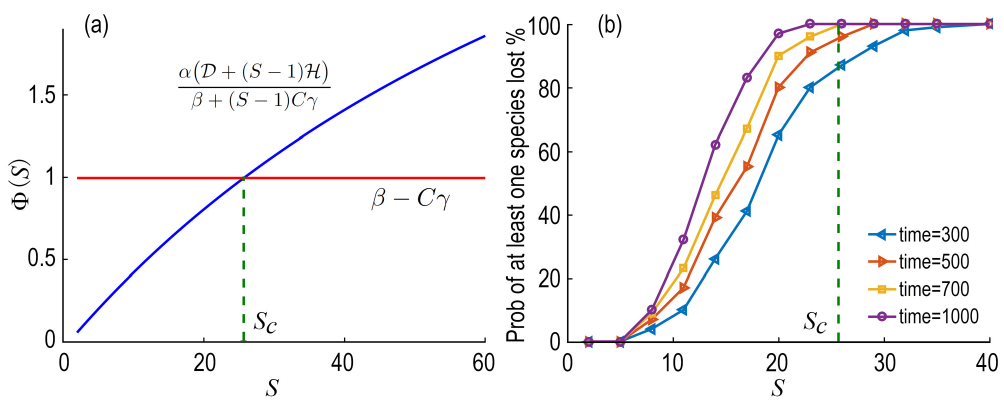}
    \end{center}
    \caption{\emph{\textbf{The critical phase transition in stochastic stability of $\mathbf{n}^*$ driven by species diversity ($S$).}} (\textbf{a}) The intersection of $\Phi(S)=\frac{\alpha \big (\mathcal{D}+(S-1) \mathcal{H} \big )}{\beta +(S-1) C \gamma}$ (blue curve) and $\beta-C \gamma$ (red line) corresponds to the critical threshold $S_c$. (\textbf{b}) The stochastic simulation results for the probability that at least one species disappears in the system in some different time periods. The parameters are taken as $\alpha=10$, $\beta=1$, $\gamma=0.01$, $\mathcal{D}=0$ and $\mathcal{Q}=0.005$, where the critical threshold $S_c$ is given by $S_c=25.69$. The probability is estimated from $100$ independent simulation runs.}
    \label{Fig2}
\end{figure}

\emph{\textbf{Discussion.}}  In this study, by analyzing a simple stochastic LV equation, we argue that the two seemingly conflicting views within the diversity-stability debate should not merely be interpreted as different ways of measuring community stability. Instead, they arise from a dimensionality-induced critical phase transition in the stochastic stability of communities; that is, a transition separating the regime dominated by statistical averaging and a systemic tipping point. Therefore, the temporal variability severs as a metric of community stability only if a stationary distribution exists around the point $\mathbf{n}^*$. Once $S$ exceeds the critical threshold $S_c$, the point $\mathbf{n}^*$ not only loses its stochastic stability, but the stationary distribution around $\mathbf{n}^*$ also disappears. We here have to point out that our model is highly simplified, but it still has significant theoretical value for revealing the underlying logic behind the diversity-stability debate.

The species asynchrony \cite{yachi1999,Loreau2008,white2023,liang2025}, that is, different species respond to environmental fluctuations in different ways, is considered as one of the main reasons why increasing diversity reduces community's temporal variability \cite{yachi1999,white2023}. For example, both the `insurance hypothesis' \cite{yachi1999} and `statistical averaging effect' \cite{doak1998,lhomme2002, zhao2022} should involve species responding relatively independently to environmental fluctuations. In our model, the effects of environmental noise on the intraspecific interactions and on the interspecific interactions are all assumed to be independent of each other. This assumption is not only similar to May's model \cite{may1972} but also consistent with the species asynchrony mechanism. However, we also found that the difference between how environmental noise affects intraspecific interactions and how it affects interspecific interactions might have an important impact on the relationship between diversity and community's temporal variability. For example, our results suggest that if the impact of environmental noise on intraspecific interactions is much weaker than on interspecific interactions, then when the diversity level of a community is very low, increasing diversity might also lead to an increase in the temporal variability. Of course, this phenomenon is only likely to occur when the community's diversity level is very low.

Finally, for the stochastic stability of the equilibrium $\mathbf{n}^*$, unlike May \cite{may1972} and some prior works \cite{theo2018, allesina2012} that primarily analyze the eigenvalue distribution of the random community matrix at $\mathbf{n}^*$, we focus on the stochastic stability of the zero solution for the It$\hat{\text{o}}$ SDE corresponding to the random community matrix, and derive the transition threshold $S_c$ governing phase shifts in stochastic stability of $\mathbf{n}^*$, where the transition is triggered by dimensionality variation (i.e. changes in $S$). Furthermore, the stochastic stability of a community in a stochastic environment inevitably involves the mechanisms of species coexistence. While many factors govern the stable coexistence  \cite{dong2015,arnoldi2019,chen2025,lechon2026}, our model demonstrates that environmental stochasticity imposes an intrinsic upper bound on community diversity. This mechanism may provide a general explanation for why community diversity cannot increase indefinitely.

\section*{Acknowledgements}

\textbf{Funding:} This work was supported by the National Nature Science Foundation of China (Grants No.32471553, No.32425036, No.32271553 and No.32271554).

\bibliographystyle{unsrt}

\newpage

\setcounter{equation}{0}
\renewcommand{\theequation}{S\arabic{equation}}

\section*{Supporting Information}

\subsection*{Stability analysis of Eq. (3)}

Note that the equilibrium of Eq. (3) is
\begin{align}
n^*_i &= n^* = \frac{\alpha}{\beta +(S-1) C \gamma}
\end{align}
for $i=1,2,\cdots,S$, and the Jacobian matrix about $\mathbf{n}^*$ can be expressed as
\begin{align}
n^* (-\beta +C \gamma) \ \mathbf{I} -n^* C \gamma \ \mathbf{J} \ ,
\end{align}
where $\mathbf{J}$ is all-ones matrix. The eigenvalues of this matrix can be given by
\begin{align}
\lambda_1 &= -n^* \big ( \beta +(S-1) C \gamma \big ) < 0 \ , \nonumber \\
\lambda_i &= n^* (-\beta +C \gamma )
\end{align}
for $i=2,3,\cdots,S$. Therefore, the equilibrium $\mathbf{n}^*$ is asymptotically stable if and only if $\beta -C \gamma >0$.

\subsection*{Analysis of the steady-state statistics}

Based on the boundary conditions of the Fokker-Planck equation in Eq. (5), which are $\lim \limits_{x_i \rightarrow \pm \infty} p(\mathbf{x},t) = 0$ and $\lim \limits_{x_i \rightarrow \pm \infty} \partial p(\mathbf{x},t) \big / \partial x_i = 0$ for $i=1,2, \cdots, S$ \cite{risken1992,kampen1997}, we have that
\begin{align}
\frac{d \left <x_i \right >}{dt} &= -n^* \Big (\beta \left <x_i \right > + \sum_{j \ne i} C \gamma \left < x_j \right > \Big ) < 0
\end{align}
for $i=1, 2, \cdots, S$ (that is, $\left < x_i \right >$ monotonically approaches zero),
\begin{align}
\frac{d \left < x_i^2 \right >}{dt} &= - 2n^* \Big ( \beta \left < x_i^2 \right > + \sum_{j \ne i} C \gamma \left < x_i x_j \right > \Big ) +2 (n^*)^4 \Big (\mathcal{D} + \sum_{j \ne i} \mathcal{H} \Big )
\end{align}
for $i=1, 2, \cdots, S$, and
\begin{align}
\frac{d \left < x_k x_l \right >}{dt} &= -n^* \Big ( \beta \left < x_l x_k \right > + \sum_{j \ne k} C \gamma \left < x_l x_j \right > \Big ) -n^* \Big ( \beta \left < x_k x_l \right > + \sum_{j \ne l} C \gamma \left < x_k x_j \right > \Big )
\end{align}
for all possible $k \ne l$, where $\left < x_i^2 \right >$ is the variance of species $i$ and $\left < x_k x_l \right >$ the covariance between species $k$ and species $l$ \cite{risken1992,kampen1997}.

Under the definition of the stochastic LV equation in Eq. (1), all species should have the same expected size and the same variance, and the covariance between different species should also be the same. Therefore, the steady solutions of Eqs. (17, 18) can be given by
\begin{align}
\left < x_i^2 \right > &= \frac{\alpha^3 \big (\mathcal{D}+(S-1) \mathcal{H} \big ) \big ( \beta + (S-2) C \gamma \big )}{( \beta -C \gamma) \big ( \beta + (S-1) C \gamma  \big )^4}
\end{align}
for all $i=1,2,\cdots,S$, and
\begin{align}
\left <x_k x_l \right > &= - \frac{\alpha^3 \big (\mathcal{D} +(S-1) \mathcal{H} \big ) C \gamma}{(\beta -C \gamma) \big ( \beta + (S-1) C \gamma \big )^4} < 0
\end{align}
for all possible $k \ne l$.

\subsection*{Stochastic stability analysis of Eq. (10)}

For convenience, let $\mathbf{E}^{(i,j)}$ be a matrix unit with entries $e_{ij}^{(i,j)}=1$ and $e_{kl}^{(i,j)}=0$ for all $(k,l) \ne (i,j)$. Thus, we have $\mathbf{B}^{(i,i)} = -n^* \sqrt{2\mathcal{D}} \ \mathbf{E}^{(i,i)}$ for $i=1,2,\cdots,S$ and $\mathbf{B}^{(i,j)} = -n^* \sqrt{2H} \ \mathbf{E}^{(i,j)}$ for all $i \ne j$.

For Eq. (10), take $V=|\mathbf{x} |^2$. Then, we have $LV(\mathbf{x})=\mathbf{x}^T \mathbf{M} \mathbf{x}$, in which
\begin{align}
\mathbf{M} &= \tilde{\mathbf{A}} + \tilde{\mathbf{A}}^T +\mathbf{S} \ ,
\end{align}
where
\begin{align}
\mathbf{S} &= \sum_{i=1}^S \sum_{j=1}^S \left ( \mathbf{B}^{(i,j)} \right )^T \mathbf{B}^{(i,j)} \nonumber \\
&= \sum_{i=1}^S \sum_{j=1}^S \mathbf{B}^{(j,i)} \mathbf{B}^{(i,j)} \nonumber \\
&= \sum_{i=1}^S \sum_{j=1}^S \left (b_{ij}^{(i,j)} \right )^2 \mathbf{E}^{(j,j)} \nonumber \\
&= \Big [ (S-1) \big (-n^* \sqrt{2\mathcal{H}} \big )^2 +\big (-n^*\sqrt{2\mathcal{D}} \big )^2 \Big ] \ \mathbf{I} \nonumber \\
&= 2 (n^*)^2 \Big [ \mathcal{D} +(S-1) \mathcal{H} \Big ] \ \mathbf{I} \ .
\end{align}
From the stochastic stability theory of SDE \cite{mao1994, mao2007}, the zero solution of Eq. (10), $\mathbf{x}=\mathbf{0}$, is stochastically stable, or mean-square exponentially stable, if and only if the matrix $\mathbf{M}$ is negative definite.

Note that
\begin{align}
m_{ii} &= -2n^* \beta + 2 (n^*)^2 \big [ \mathcal{D} +(S-1) \mathcal{H} \big ]
\end{align}
for $i=1,2,\cdots,S$, and
\begin{align}
m_{ij} &= m_{ji} = \tilde{a}_{ij} + \tilde{a}_{ji} = -2n^* C \gamma
\end{align}
for all $i \ne j$. Thus, $\mathbf{M}$ can be re-expressed as
\begin{align}
\mathbf{M} &= \Big [ -2n^* \beta +2(n^*)^2 \big (\mathcal{D} +(S-1) \mathcal{H} \big ) +2n^* C \gamma \Big ] \ \mathbf{I} -2n^*C \gamma \ \mathbf{J} \ ,
\end{align}
where $\mathbf{J}$ is all-ones matrix. The eigenvalues of $\mathbf{M}$ are given by
\begin{align}
\lambda_1 &= -2n^* \beta +2 (n^*)^2 \big (\mathcal{D} +(S-1)\mathcal{H} \big ) +2n^* C \gamma -2n^* C \gamma S \nonumber \\
&= -2n^* \beta + 2 (n^*)^2 \big (\mathcal{D} +(S-1)\mathcal{H} \big ) -2n^* (S-1) C \gamma \nonumber \\
&= 2n^* \Big [ -\beta +n^* \big (\mathcal{D} +(S-1) \mathcal{H} \big ) -(S-1) C \gamma \Big ] \ , \nonumber \\
\lambda_i &= -2n^* \beta +2 (n^*)^2 \big (\mathcal{D} +(S-1)\mathcal{H} \big ) +2n^* C \gamma \nonumber \\
&= 2n^* \Big [ -\beta +n^* \big (\mathcal{D} +(S-1) \mathcal{H} \big ) + C \gamma \Big ]
\end{align}
for $i=2,3,\cdots,S$. Therefore, the matrix $\mathbf{M}$ is negative definite if and only if
\begin{align}
& \ n^* \big (D +(S-1) H \big ) < \beta -C \gamma \nonumber \\
\Rightarrow & \ \ \ \frac{\alpha \big (D +(S-1) H \big )}{\beta +(S-1) C \gamma} < \beta -C \gamma \ .
\end{align}

\end{document}